\documentclass[twocolumn,twocolappendix]{aastex631}

\shorttitle{A Star-based Method for Precise Flux Calibration of the CSST}
\shortauthors{Yang et al.}

\usepackage{subfigure}
\usepackage{amsmath}
\usepackage{threeparttable}

\makeatletter
\newcommand{\rmnum}[1]{\romannumeral #1}
\newcommand{\Rmnum}[1]{\expandafter\@slowromancap\romannumeral #1@}
\makeatother

\begin{document}

\title{A star-based method for precise flux calibration of the Chinese Space Station Telescope (CSST) slitless spectroscopic survey}

\correspondingauthor{Haibo Yuan}
\email{yuanhb@bnu.edu.cn}

\author[0000-0002-9824-0461]{Lin Yang}
\affiliation{Department of Cyber Security, Beijing Electronic Science and Technology Institute, Beijing, 100070, China}
\affiliation{College of Artificial Intelligence, Beijing Normal University No.19, Xinjiekouwai St, Haidian District, Beijing, 100875, P.R.China}

\author[0000-0003-2471-2363]{Haibo Yuan}
\affiliation{Institute for Frontiers in Astronomy and Astrophysics, Beijing Normal University, Beijing, 102206, China}
\affiliation{Department of Astronomy, Beijing Normal University, Beijing, 100875, People's Republic of China}

\author[0000-0002-3849-8532]{Fuqing Duan}
\affiliation{College of Artificial Intelligence, Beijing Normal University No.19, Xinjiekouwai St, Haidian District, Beijing, 100875, P.R.China}

\author[0000-0003-1863-1268]{Ruoyi Zhang}
\affiliation{Institute for Frontiers in Astronomy and Astrophysics, Beijing Normal University, Beijing, 102206, China}
\affiliation{Department of Astronomy, Beijing Normal University, Beijing, 100875, People's Republic of China}

\author[0000-0002-1259-0517]{Bowen Huang}
\affiliation{Institute for Frontiers in Astronomy and Astrophysics, Beijing Normal University, Beijing, 102206, China}
\affiliation{Department of Astronomy, Beijing Normal University, Beijing, 100875, People's Republic of China}

\author[0000-0001-8424-1079]{Kai Xiao}
\affiliation{School of Astronomy and Space Science, University of Chinese Academy of Sciences, Beijing 100049, People's Republic of China}

\author[0000-0003-3535-504X]{Shuai Xu}
\affiliation{Institute for Frontiers in Astronomy and Astrophysics, Beijing Normal University, Beijing, 102206, China}
\affiliation{Department of Astronomy, Beijing Normal University, Beijing, 100875, People's Republic of China}

\author[0009-0005-7743-6229]{Jinming Zhang}
\affiliation{Institute for Frontiers in Astronomy and Astrophysics, Beijing Normal University, Beijing, 102206, China}
\affiliation{Department of Astronomy, Beijing Normal University, Beijing, 100875, People's Republic of China}

\begin{abstract}

The upcoming Chinese Space Station Telescope (CSST) slitless spectroscopic survey poses a challenge of flux calibration, which requires a large number of flux-standard stars. In this work, we design an uncertainty-aware residual attention network, the UaRA-net, to derive the CSST SEDs with a resolution of $R=200$ over the wavelength range of 2500--10000\,$\rm \AA$ using LAMOST normalized spectra with a resolution of $R=2000$ over the wavelength range of 4000--7000\,$\rm \AA$. With the special structure and training strategy, the proposed model can not only provide accurate predictions of SEDs but also their corresponding errors. The precision of the predicted SEDs depends on effective temperature (Teff), wavelength, and the LAMOST spectral signal-to-noise ratios (SNRs), particularly in the GU band. For stars with Teff = 6000 K, the typical SED precisions in the GU band are 4.2\%, 2.1\%, and 1.5\% at SNR values of 20, 40, and 80, respectively. As Teff increases to 8000 K, the precision increases to 1.2\%, 0.6\%, and 0.5\%, respectively. The precision is higher at redder wavelengths. In the GI band, the typical SED precisions for stars with Teff = 6000 K increase to 0.3\%, 0.1\%, and 0.1\% at SNR values of 20, 40, and 80, respectively. We further verify our model using the empirical spectra of the MILES and find good performance. The proposed 
method will open up new possibilities for optimal utilization of slitless spectra of the CSST and other surveys.

\end{abstract}


\section{Introduction} \label{introduction}

Slitless spectroscopy is a powerful tool in astronomy, with the capability of obtaining spectra of huge numbers of objects that are free from target selection biases. Ground-based slitless spectroscopic surveys have a long history, such as the Curtis Schmidt - thin prism survey \citep{MacAlpine1977}, the Second Spectral Survey \citep{Markarian1987}, the APM QSO Survey \citep{Foltz1989}, the Hamburg Quasar Survey\citep{Hagen1995}, and the Quasars near Quasars survey \citep{Worseck2008}.
However, these ground-based slitless spectroscopic surveys are susceptible to serious problems including  source contamination and bright sky background. 
Due to the high spatial resolution and low sky background, space is the most natural and suitable place for slitless spectroscopic observations. Wide-field slitless spectroscopic surveys of space telescopes, such as the China Space Station Telescope \citep[CSST;][]{Zhan2011},
the Nancy Grace Roman Space Telescope \citep[][]{Green2012,Akeson2019}, and the Euclid Space Telescope \citep{Laureijs2011,Laureijs2012,Amendola2013,Amendola2018}, can go much fainter and wider and thereby provide huge opportunities in astronomy.  

CSST is an up-coming 2-meter
space telescope with a large field of view (FoV) of 1.1 $\rm deg^{2}$. In ten years starting from 2025, the telescope will simultaneously conduct both imaging and slitless grating spectroscopic surveys, covering a large sky area of 17,500 $\rm deg^{2}$  and at a high spatial resolution of $\sim 0.15$ arcsec  \citep{Zhan2011,Cao2018,Gong2019}. With seven photometric filters (NUV, u, g, r, i, z, and y), the imaging survey spans a broad spectrum of wavelengths from 255 nm to 1000 nm. The slitless spectroscopic survey complements the imaging survey by offering high quality spectra for hundreds of millions of stars and galaxies in three bands, GU (255–420 nm), GV (400–650 nm), and GI (620–1000 nm), within the same wavelength range. The spectra have a resolution higher than 200 and a magnitude limit around 23 mag in the GU/GV/GI band.
The designed parameters and intrinsic transmission curves for the CSST photometric and spectroscopic surveys can be found in Table 1 and Figure 1 of \cite{Gong2019}, respectively.

To make optimal usage of the unique capability of wide-field slitless spectroscopic surveys,  data reduction 
is very important and requires a lot of efforts. In the data reduction process,  wavelength and flux calibrations are extremely challenging, particularly for the CSST. 

The challenges of wavelength calibration for the CSST were discussed in \cite{Yuan2021}.  Briefly speaking, there are not sufficient number of specific emission line objects and sufficient calibration time to eliminate field effects and geometric distortions. To solve this problem, \cite{Yuan2021} proposed a star-based method for the wavelength calibration of the CSST slitless spectroscopic survey. This method makes use of three prerequisites: (\rmnum{1}) Over ten million stars with precise radial velocities have been delivered by spectroscopic surveys like LAMOST \citep{Deng2012,Liu2014}; (\rmnum{2}) The CSST can observe a large number of such stars during a short time thanks to its large FoV; (\rmnum{3}) A narrow segment of CSST spectra can provide reliable estimates of radial velocities \citep{Sun2021}.
Taking advantages of the above prerequisites, the key idea of this method is to use enormous numbers of stars (absorption lines rather than emission lines) of known radial velocities observed during normal scientific observations as wavelength standards to monitor and correct for possible errors in wavelength calibration.
Using only hundreds of velocity standard stars, they demonstrated that it is possible to achieved the wavelength calibration precision of a few km $\rm s^{-1}$ for the GU band, and about 10 to 20 km $\rm s^{-1}$ for the GV and GI bands.

In this work, we focus on achieving precise flux calibration for the CSST slitless spectra. The goal of flux calibration is to convert astronomical measurements from instrumental units to physical units. 
The challenges of flux calibration for slitless spectroscopic observations lie in 
flat-fielding. A 3D flat-field cube must be used to perform flat-field correction for each pixel in the 2D detector as a function of incident wavelength.
To characterize the wavelength-dependent behavior, a series of direct-image flat fields taken at different central wavelengths can be used to construct the 3D flat-field cube for telescopes such as 
the Hubble Space Telescope \citep[HST;][]{Momcheva2016,Pharo2020}, 
the Roman Space Telescope, and the Euclid Space Telescope. Contrasting with the conventional approach that relies on a filter wheel to switch between filters and gratings, the CSST employs a strategy in which each detector is dedicated to a specific filter or grating (see Figure 5 of \cite{Zhan2021}). As a result, this hardware design makes direct imaging observations infeasible for the CSST slitless spectroscopic survey.

To perform high-precision photometric calibration of wide-field imaging surveys, 
\citet{Yuan2015} proposed a stellar color regression (SCR) method in the era of 
large-scale spectroscopic surveys. 
The method assumes that stellar intrinsic colors can be precisely predicted by their atmospheric parameters, thus millions of stars targeted by spectroscopic surveys like LAMOST can serve as excellent color standards. Combining uniform photometric data from Gaia, the method has been applied to 
a number of surveys, including SDSS/Stripe 82 \citep{Huang2022a}, SMSS DR2 \citep{Huang2021}, PS1 DR1 \citep{Xiao2022,Xiao2023}, Gaia DR2 \citep{Niu2021a} and EDR3 \citep{Niu2021b, Yang2021}. A typical precision of a few mmag was usually achieved. See a recent review by \citet{Huang2022b}.

Assuming that stars with the same normalized spectra have the same spectral energy distributions (SEDs), in this work, 
we further develop the SCR method to predict the CSST-like ($R \sim 200$) SEDs 
for a huge number of stars from LAMOST-like ($R \sim 1800$) normalized spectra. Then, with a large number of flux standard stars,
we can map the 2D variations of the large-scale flat\footnote{The small-scale flats can be obtained for a small number (around 12) of wavelengths covering from the near-UV to the near-IR by the LED light on board. Therefore, small-scale flats are ignored in this work.} in the detector for each wavelength. Combining the 2D large-scale flats together, we can obtain the 3D flat-field cube for accurate flux calibration of the CSST slitless spectroscopic survey.

We use a neural network to predict the 
CSST-like SEDs from LAMOST-like normalized spectra.
Due to the impact of observational noise on model performance, it is challenging to deal with uncertain data. Traditional denoising methods depend on user-defined filters \citep[e.g.][]{Gilda2019,Politsch2020}. Without manual denoising, various deep learning methods have been put forward to automatically learn important features from noisy observations \citep[e.g.][]{Zhao2019,Zhou2021,Zhou2022}. However, these methods depend on sufficient training sample of high SNRs and have difficulty in predicting flux errors. To address the above limitations, we propose an uncertainty-aware residual attention network, UaRA-net, to predict not only precise and robust CSST-like SEDs, but also their corresponding errors, using LAMOST-like normalized spectra and their errors as input.
Our model structure and training strategy make full use of the prior observational errors, thereby improving the flexibility on input spectra at different SNRs.

The paper is organized as follows: Section\,\ref{data} describes the data used in this work. Section \,\ref{method} introduces the proposed UaRA-net in detail. Section \,\ref{Result} reports the results of the pre-trained and fine-tuned models, as well as a verification of the proposed method. We summarize in Section \,\ref{summary} and  discuss the challenges and potential future improvements.

\section{Data} \label{data}
The datasets used in this work consist of theoretical spectra and empirical spectra. The former, for which we adopt the BOSZ models, are used to pre-train the UaRA-net model. The latter, including the NGSL and MILES spectra, are used to further fine-tune and verify the model. 

\subsection{Theoretical spectra} 
The BOSZ model atmosphere grids \citep{Meszaros2012,Bohlin2017} were constructed using ATLAS9, and provide 10 resolution modes in the R$=$200--300000 range. The BOSZ spectra\footnote{\url{https:// archive.stsci.edu/prepds/bosz/}} have a wide wavelength range of $0.1$--$32\mu \rm m$, which covers the CSST GU, GV, and GI bands well. The atmospheric parameters of the BOSZ spectra cover $T_{\rm eff}$ from 3500 to 30000 K, $\log g$ from 0.0 to 5.0, $\rm [M/H]$ from $-$2.5 to +0.5, $\rm [C/H]$ from $-$0.25 to +0.5, and $\rm [\alpha/M]$ from $-$0.75 to +0.5, with steps of 250--1000 K, 0.5, 0.25, 0.25, and 0.25 dex, respectively. 

During the pre-training stage of UaRA-net, two sets of spectra, one set with $R=2000$ in the wavelength range of 4000--7000 $\rm \AA$ and the other set with $R=200$ in the wavelength range of 2500--10500 $\rm \AA$, are chosen to simulate the LAMOST low-resolution spectra and CSST slitless spectra, respectively. 
The LAMOST-like spectra serve as the inputs of the UaRA-net and are normalized by diving their continua, which are estimated using the moving average method with a window size of 51 pixels. In cases where there are fewer than 25 pixels on one side of the boundary, all pixels on that side are used.
The CSST-like spectra are used as the outputs of the UaRA-net and are scaled by dividing the mean flux value in the wavelength range of 5526--5586 $\rm \AA$. The above wavelength range is adopted because it is free of strong stellar absorption features and close to the central wavelength of the V band.

Only stars or spectra with the $T_{\rm eff}$ ranging from 5000 to 9750 K are used. Hotter stars are excluded, because there will be limited number of such stars in the CSST data; cooler stars are excluded either, as their scaled fluxes in the GU band are very low and difficult to predict. The selected stars are randomly divided into training and testing samples in the ratio of 3:1. The training set contains 41219 stars. For each training star, we add noise to its clean LAMOST-like spectrum to generate spectra at different SNRs $\in \{20, 25, 30, 35, 40\}$, resulting in 206095 training noisy spectra. The testing set contains the other 13740 stars, and their clean LAMOST-like spectra are mixed with noise at different SNRs $\in \{20, 25, 30, 35, 40, 80\}$, resulting in 82440 testing noisy spectra. It should be noted that the testing set also includes spectra with SNR=80, which are used to examine the extrapolation ability of the proposed method to high-quality spectra. 
A noisy spectrum $F$ with SNR $=snr$ at wavelength $\lambda$ can be written as 
\begin{equation}\label{multiplicative noise}
  {F_\lambda}'=F_\lambda+r*F_\lambda, 
\end{equation}
where $r\sim\mathcal{N}(0,(1/snr)^2)$.

\subsection{Empirical spectra} 
The MILES spectral stellar library\footnote{\url{http://research.iac.es/proyecto/miles/pages/stellar-libraries/miles-library.php}} consists of 985 flux-calibrated and reddening-corrected spectra observed by the 2.5m INT telescope, covering the wavelength range of 3500--7500$\rm \AA$ \citep{Sanchez-Blazquez2006} at a spectral resolution of $R=2500$ \citep{Falcon-Barroso2011}. The MILES catalogue \citep{Cenarro2007} provides stellar effective temperatures ($T_{\rm eff}$), surface gravities ($\log g$), and metallicities ([Fe/H]). Similar to the BOSZ spectra, we require stars in the same $T_{\rm eff}$ range of 5000--9750 K. To ensure the quality of the MILES spectra, we further require stars with $E(B-V)<0.1$ mag. To exclude abnormal MILES spectra (see more details in Appendix\,\ref{data selection}), we train a Gaussian process regression (GPR) model to predict SEDs at $R=200$ from stellar parameters ($T_{\rm eff}$, $\log g$, and [Fe/H]).
In this paper, relative residual of $r$ with $n$ pixels is used to represent the difference between the predicted SEDs and the true SEDs, which is defined as
\begin{equation}\label{F_err}
    r=\frac{\hat F-F}{F}
\end{equation}
where $\hat F$ is a n-dimensional vector that represents the predictive SED. 
Then, the root mean square relative error ($RMSAE$) that represents the overall difference between the predicted SEDs and the true SEDs is defined as
\begin{equation}\label{RMSAE}
    RMSAE=\sqrt{\frac{1}{n}\sum_{n}^{i=1}r^2}
\end{equation}
We regard MILES spectra of $RMSAE<0.05$ as normal, and obtain 265 stars.

The NGSL spectral stellar library\footnote{\url{https://archive.stsci.edu/pub/hlsp/stisngsl/v2/stis_ngsl_v2.zip}} comprises 378 stars observed by the Hubble Space Telescope Imaging Spectrograph using the three low dispersion ($R\sim 1000$) gratings, G230LB, G430L, and G750L, which overlap at 2990--3060$\rm \AA$ and 5500--5650$\rm \AA$. The spectra cover the wavelength range of 1670--10250$\rm \AA$.
Reddening corrections and flux corrections were applied to their SEDs, and their zero point is the same as the BOSZ spectra, using the provided reddening values $A_V$ and calibration curves calculated by the difference from the CALSPEC reddening spectra of seven common stars. To avoid spectra overlapping effect, the calibration curve is only applied to the NGSL SED segment at $\lambda> 5700\rm \AA$. 
Similar to the BOSZ spectra, we also require stars with $T_{\rm eff}$ in the range of 5000–9750 K. To ensure the quality of the NGSL spectra, only stars with $-0.05<A_V<0.25$ mag, err($\log g$, dex) $<$ 0.7 dex, and err([Fe/H]) $<$ 0.25 are selected. Before using NGSL spectra to fine-tune the UaRA-net, we train two GPR models, one to exclude abnormal spectra (see more details in Appendix\,\ref{data selection}), and the other to generate normalized spectra at $R=2000$ (see more details in Appendix\,\ref{NGSL}) from stellar parameters ($T_{\rm eff}$, $\log g$, and [Fe/H]). The first model takes stellar parameters as inputs and returns predicted SEDs at $R=200$. By comparing the difference between the predicted SEDs and the ture SEDs, NGSL spectra of $RMSAE<0.05$ are regared as normal and 104 stars are selected. The second model is trained on the MILES spectra and applied on the NGSL spectra to generate normalized NGSL-based spectra at $R=2000$. During the fine-tuning stage of UaRA-net, 104 generated NGSL-based spectra are randomly divided into training and testing samples in a ratio of 4:1.

During the fine-tuning stage of UaRA-net, the  generated NGSL-based spectra are used as inputs and the scaled NGSL observations degraded to $R=200$ at $2500<\lambda<10250\rm \AA$ are used as outputs. During the stage of verifying the UaRA-net, the normalized MILES spectra degraded to $R=2000$ at $4000<\lambda<7000\rm \AA$ are used as inputs and the scaled MILES observations degraded to $R=200$ at $3500<\lambda<7500\rm \AA$ are used as outputs.

\section{Method} \label{method}
\begin{figure*}[ht!]
  \centering
  \includegraphics[width=\textwidth]{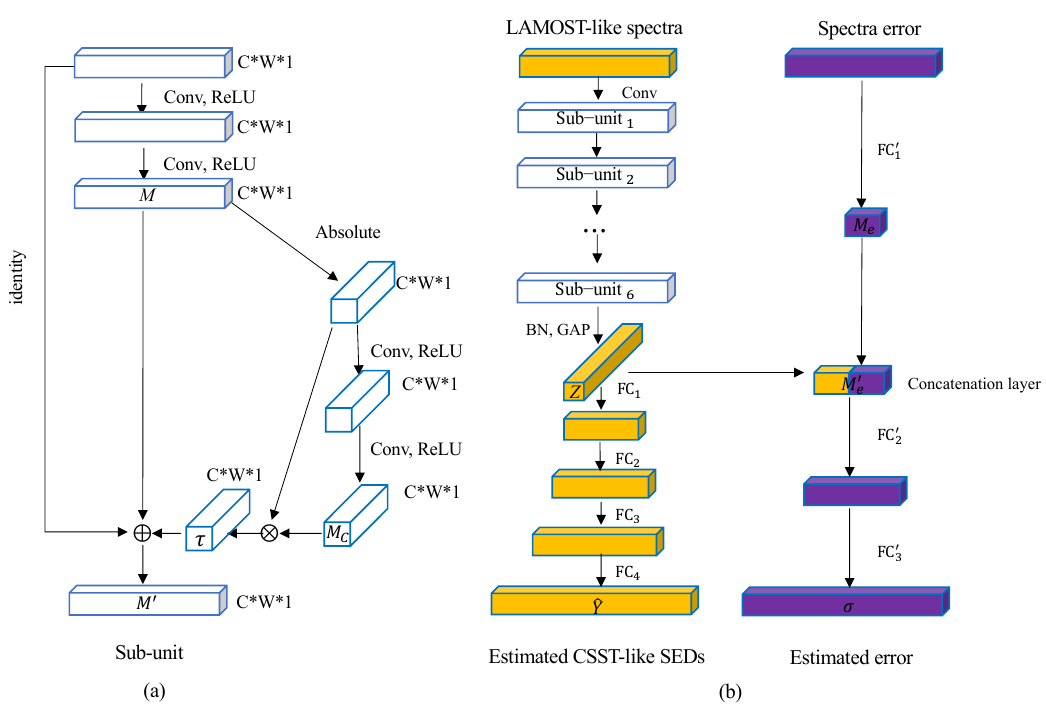}
  \caption{(a) A sub-unit of the proposed network, where $C$, $W$, and 1 in $C * W * 1$ are the indicators of the number of channels, width, and height of the feature map, respectively. (b) Overall architecture of the proposed network.}
  \label{fig:network}
\end{figure*}
Considering the distribution of the input highly noisy spectra, we propose an uncertainty-aware residual attention network (UaRA-net) that estimates CSST-like SEDs and their corresponding errors by returning a probability distribution with the mean and variance of a Gaussian. As shown in Fig.\,\ref{fig:network}, the UaRA-net is divided into two branches: the SED branch and the uncertainty branch. The SED branch takes normalized LAMOST-like spectra as the input vector $X$ and returns a latent representation $Z$ and the predicted means $\mu$ of the CSST-like SEDs. The uncertainty branch takes the concatenation of the corresponding spectra errors $X_{err}$ and $Z$ as the input vector and returns the predicted heteroscedastic uncertainties $\sigma$ of the CSST-like SEDs.

\subsection{Residual attention network} \label{SEDs estimation}
The residual network (RetNet; \citealt{he2016}) is a modularized architecture that uses stack building blocks called residual units. One of the advantages of RetNet is the shortcut connection between neighboring residual units, which helps address the degradation problem in deep networks. To maintain the same connecting shape, extra zero entries are padded when the dimensions of the next residual unit increase. The residual unit in the SEDs branch consists mainly of convolutional layers, batch normalization (BN) layers, and rectified linear unit (ReLU) activation functions. The convolutional layers greatly reduce the number of trainable parameters by sharing parameters with the convolutional kernel. The BN layers prevent overfitting and speed up convergence by ensuring non-zero variances of the forward-propagated signals.

To improve the robustness of the network regarding the noise, the attention map $M_c \in \mathbb{R}^{C*W*1}$, which is generated by three sequential 1D convolutions with ReLU activation function, first serves as the denoising operator on feature maps $M$ by determining flux-related soft thresholds $\tau$. The threshold estimates are calculated as
\begin{gather}\label{threshold}
    M_c=\Phi_{cov,Relu}(|M|)\\
    \tau=M_c\otimes |M|
\end{gather}
where $\Phi_{cov,Relu}(\bullet)$ denotes a linear combination of non-linear 1D convolution functions and $\otimes$ denotes element-wise multiplication. 
Then, an intermediate feature map $M^{'}\in \mathbb{R}^{C*W*1}$ can be refined as follows
\begin{equation}\label{denoising operation}
    M^{'}=\left\{\begin{matrix}
    {\rm sgn}(M)(|M|-\tau) &   |M|>  \tau& \\ 
    0 &  -\tau \leq |M| \leq \tau & \\
\end{matrix}\right.
\end{equation}
where ${\rm sgn}(x)=1$ if $x>0$ and ${\rm sgn}(x)=-1$ otherwise, and $\tau \geqslant 0_{C\times W\times 1}$ is the feature-wise threshold.

The latent representation $Z\in \mathbb{R}^{C*1}$ of the $M^{'}$ is believed to contain all information of stellar atmospheric parameters and elemental abundances (hereafter stellar labels) and is calculated by a global average pooling (GAP) operation as
\begin{equation}\label{Z}
    Z=GAP(M^{'})
\end{equation}
Sequential full connections (FC) following by the GAP map $Z$ into CSST-like SEDs by learning nonlinear functions as follows
\begin{equation}\label{SEDs}
    Y=\Phi_{FC}(Z)
\end{equation}
where $\Phi_{FC}(\bullet)$ denotes a linear combination of non-linear functions from full connected layers.

\subsection{Uncertainty quantification} \label{uncertainty quantification}

The uncertainty of the SED estimates in the UaRA-net comes from two sources: epistemic
uncertainty and aleatoric uncertainty. Given the large number of spectra in the training sample, the epistemic uncertainty is far smaller than the aleatoric uncertainty. In this work, we consider the aleatoric uncertainty, which depends on the observation error of the input spectra and stellar labels, as an approximation of the uncertainty of the SED estimates.

The error branch of the UaRA-net is used to estimate the aleatoric uncertainty. The feature map $M_e\in \mathbb{R}^{C*1}$ associated with the spectra noise is generated by the full connection as
\begin{equation}\label{Me}
    M_e=\Phi(F_e)
\end{equation}
where $F_e\in \mathbb{R}^{W*1}$ represents the input spectra noise.
The feature map $Z\in \mathbb{R}^{C*1}$ calculated by Eq.\,(\ref{Z}) is associated with the stellar labels. The above two feature maps are then concatenated by the concatenation layer as
\begin{equation}\label{Me}
    M_e^{'}=[Z;M_e]
\end{equation}
Then, two sequential full connections (FC) generate the input-dependent uncertainty of estimates associated with the spectra noise and stellar labels by non-linear mapping as
\begin{equation}\label{sigma}
    \sigma=\Phi_e(F_e)
\end{equation}

\subsection{Training strategy}
Gaussian maximum likelihood is used to model the loss function of the URA-net. We assume that the difference $\delta Y_i$ between the observed target $Y_i$ and the prediction $\hat{Y_i}$ follows Gaussian distributions
\begin{equation}
    r_i=\frac{\Phi(Z)-Y_i}{Y_i}
\end{equation}
where $r_i \sim \mathcal N(0,\sigma_i ^2)$ is the heteroscedastic uncertainty. 
The predictive SED in Eq.\,(\ref{SEDs}) can be defined as 
\begin{equation}
    Y_i=\Phi(Z)-r_i\cdot Y_i
\end{equation}
In the probabilistic view, we assume that the prediction SED $Y_i$ is the mean of a Gaussian predictive distribution with variance $\sigma_i Y_i$, which can be defined as
\begin{equation}
    Y_i \sim \mathcal N(\Phi(Z),(\sigma_i Y_i)^2)
\end{equation}
The probability of $Y_i$ given a spectra input $F_i$ can be approximated by a Gaussian distribution
\begin{equation}
\begin{aligned}
    p(Y_i|F_i)&=\frac{1}{\sqrt{2\pi}\sigma_i Y_i}\times {\rm exp}\left(-\frac{(Y_i-\Phi(Z))^2}{2(\sigma_i Y_i)^{2}}\right)\\
    &=\frac{1}{\sqrt{2\pi}\sigma_i Y_i}\times {\rm exp}\left(-\frac{r_i^2}{2 \sigma_i^2}\right)
\end{aligned}
\end{equation}
In maximum likelihood inference, we define the likelihood as
\begin{equation}\label{likelihood}
    p(Y_1,...,Y_k,|F)=p(Y_1|F)...p(Y_k|F)
\end{equation}
To train $\sigma$ and $Y$, we find the parameters through minimizing the following loss $\mathcal L$, which is the negative logarithm of the Eq.\,(\ref{likelihood}) likelihood
\begin{equation}\label{loss}
    \mathcal {L}=\sum_{i=1}^{n}\left(\frac{r_i^2}{2\sigma_i^2}+\log \sigma_i+\log Y_i\right)
\end{equation}

\section{Experimental results} \label{Result}

\begin{figure*}[ht!]
  \centering
  \includegraphics[width=\textwidth]{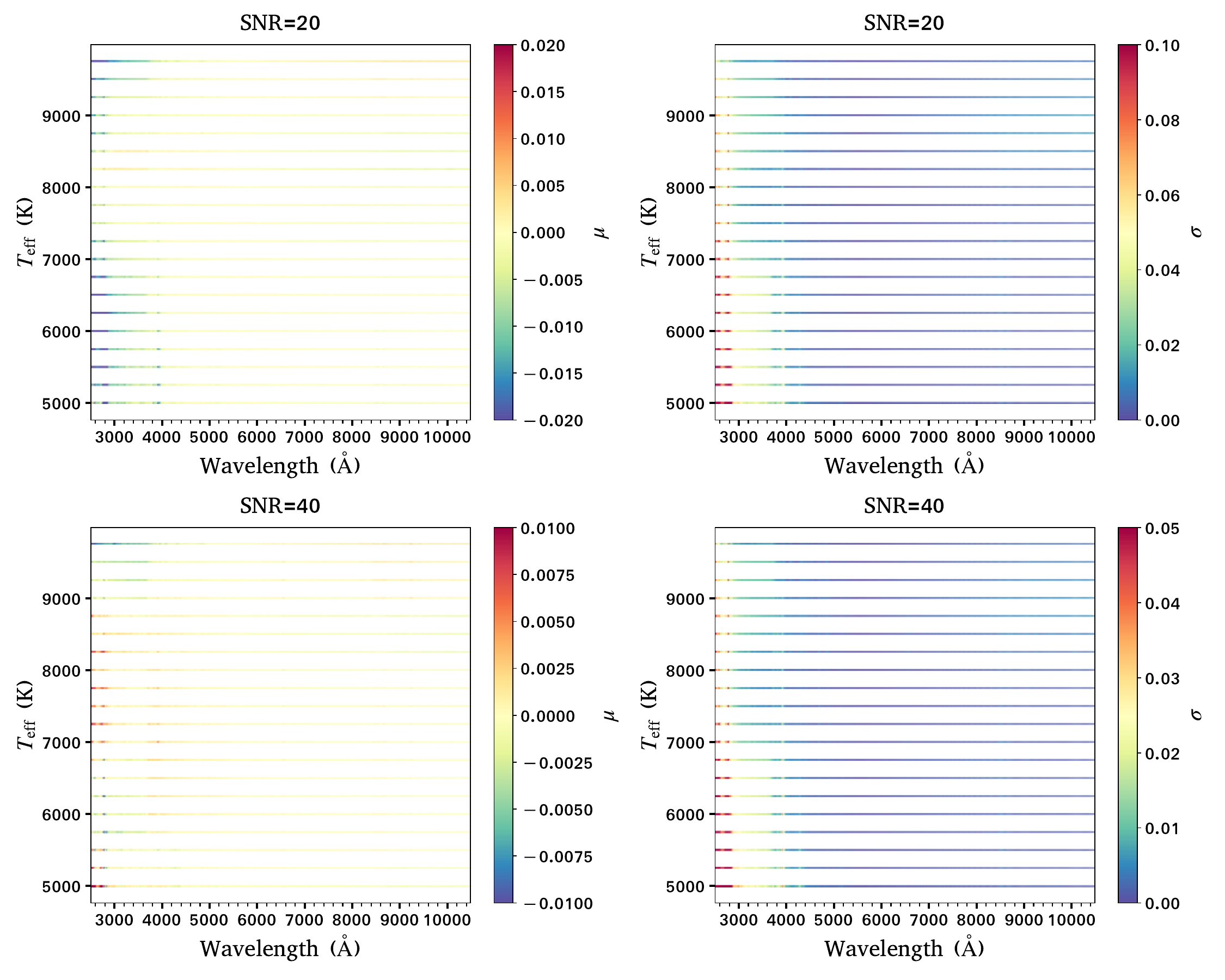}
  \caption{The medians (left column) and $1\sigma$ uncertainties (right column) of relative residuals of the training set with SNR$\in\{20,40\}$ for different $T_{\rm eff}$ bins. The relative residuals are defined in Eq.\,(\ref{F_err}).}
  \label{fig:snr_u_sigma_train}
\end{figure*}

\begin{figure*}[ht!]
  \centering
  \includegraphics[width=\textwidth]{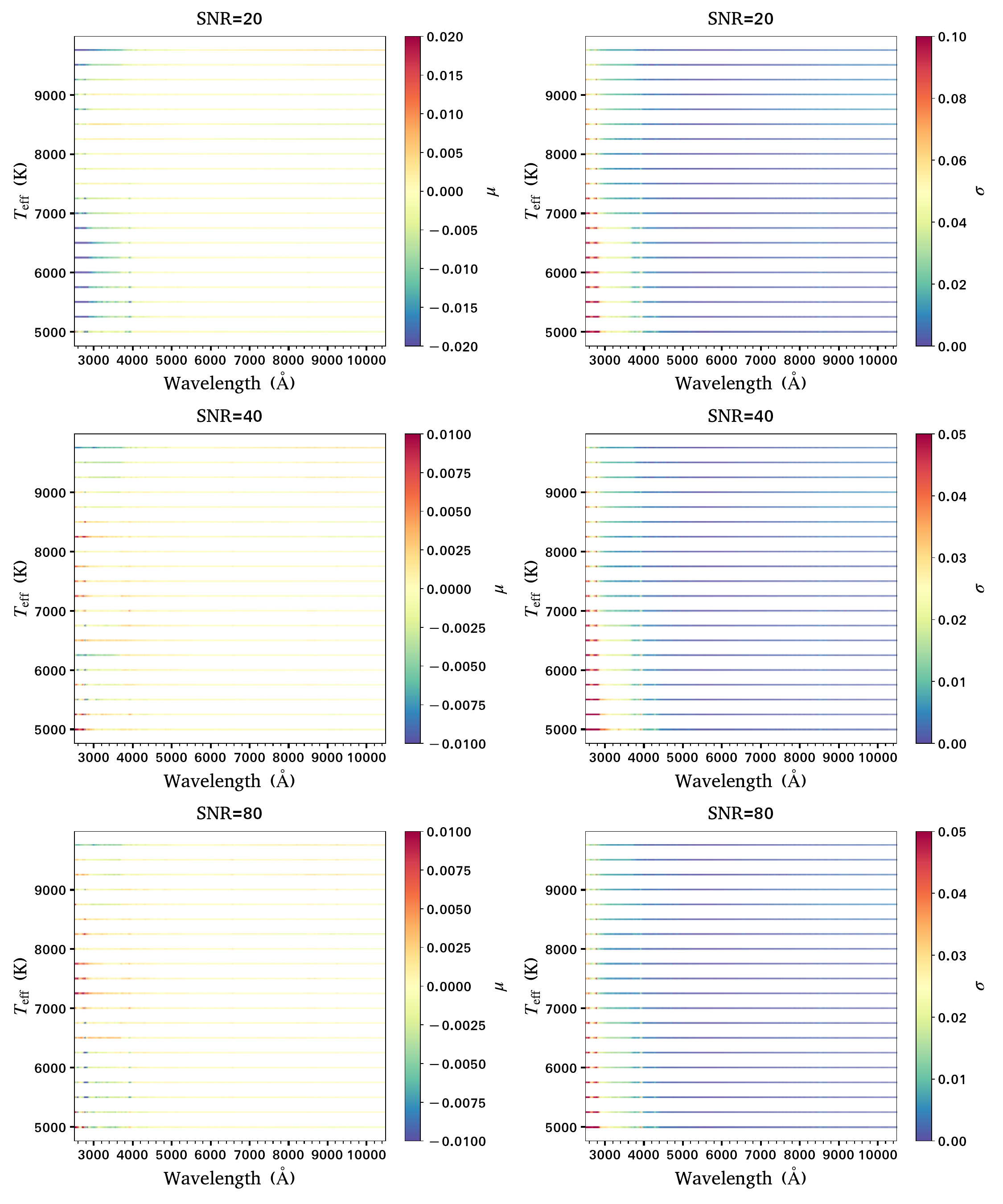}
  \caption{Similar to Fig.\,\ref{fig:snr_u_sigma_train}, but for the testing set with SNR$\in\{20,40,80\}$. It is worth mentioning that the spectra with $\rm SNR=80$ are not in the training set.}
  \label{fig:snr_u_sigma_test}
\end{figure*}

\begin{figure*}[ht!]
  \centering
  \includegraphics[width=\textwidth]{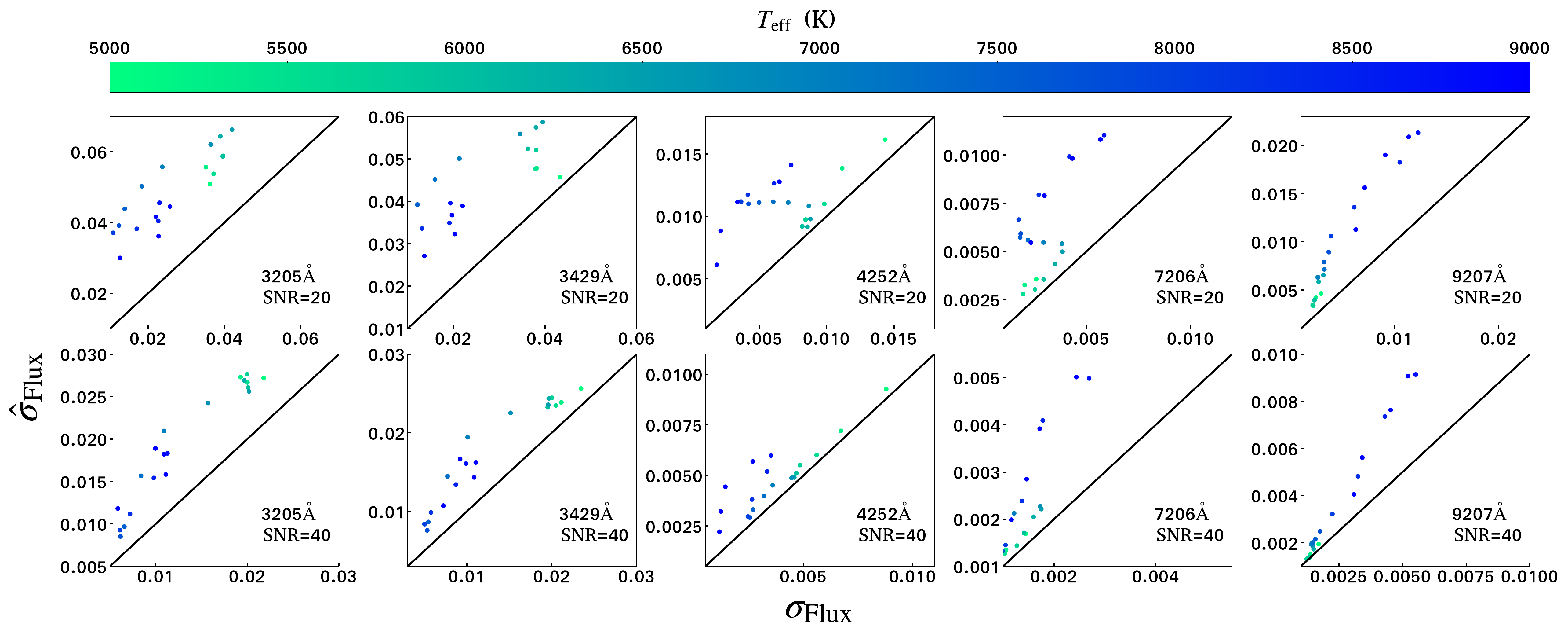}
  \caption{Comparisons between typical predicted errors (Y-axis) from the UaRA-net and $1\sigma$ uncertainties of the relative residuals (X-axis) for training set in $T_{\rm eff}$ bins with SNR$\in\{20,40\}$. To avoid crowding, five representative wavelengths at $3205\rm \AA, 3429\rm \AA, 4252\rm \AA, 7206\rm \AA, 9207\rm \AA$ are plotted.}
  \label{fig:sigma_train}
\end{figure*}

\begin{figure*}[ht!]
  \centering
  \includegraphics[width=\textwidth]{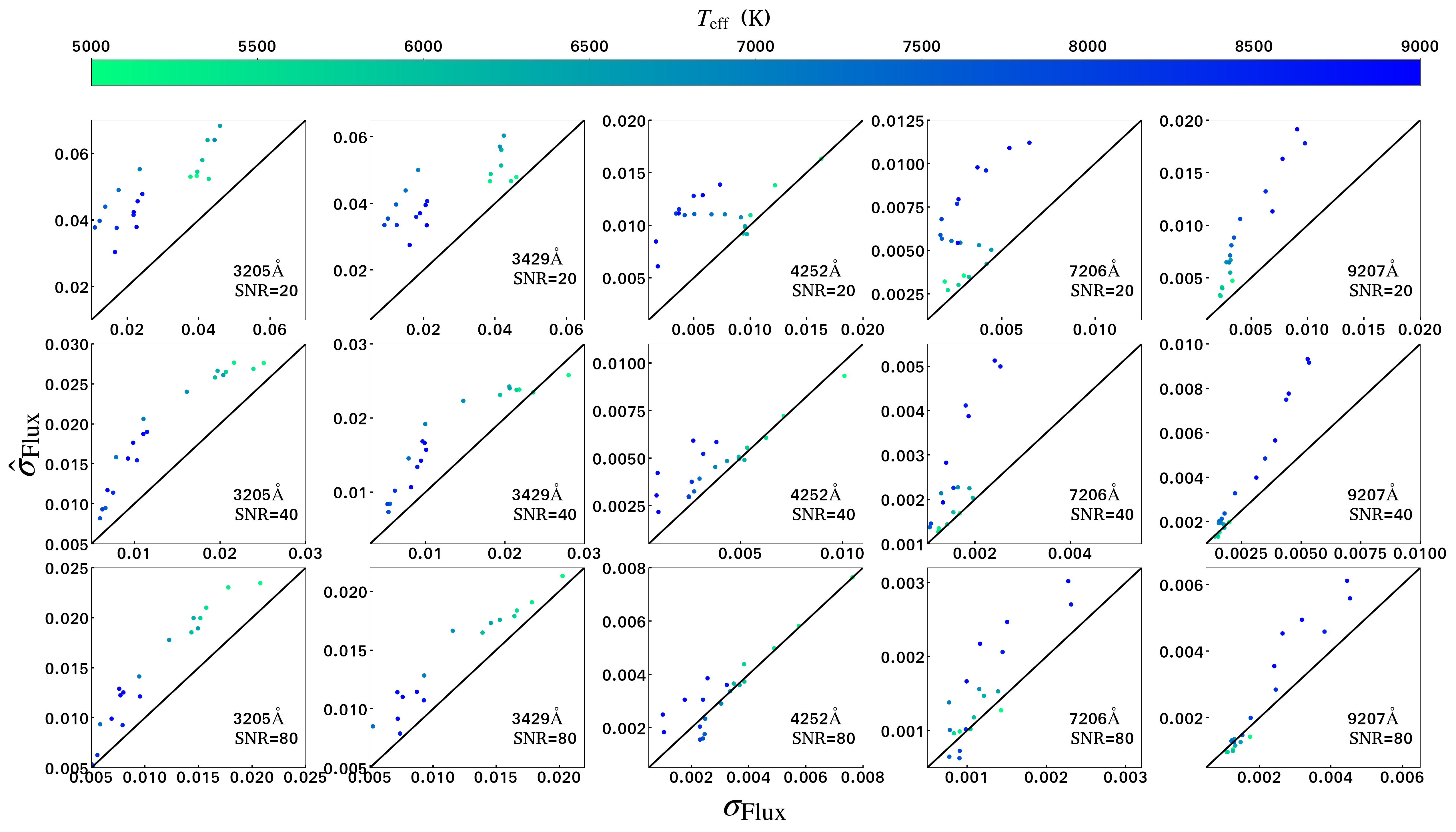}
  \caption{Similar to Fig.\,\ref{fig:sigma_train}, but for the testing set with SNR$\in\{20,40,80\}$. It is worth mentioning that the spectra with $\rm SNR=80$ are not in the training set.}
  \label{fig:sigma_test}
\end{figure*}

\begin{figure*}[ht!]
  \centering
  \includegraphics[width=\textwidth]{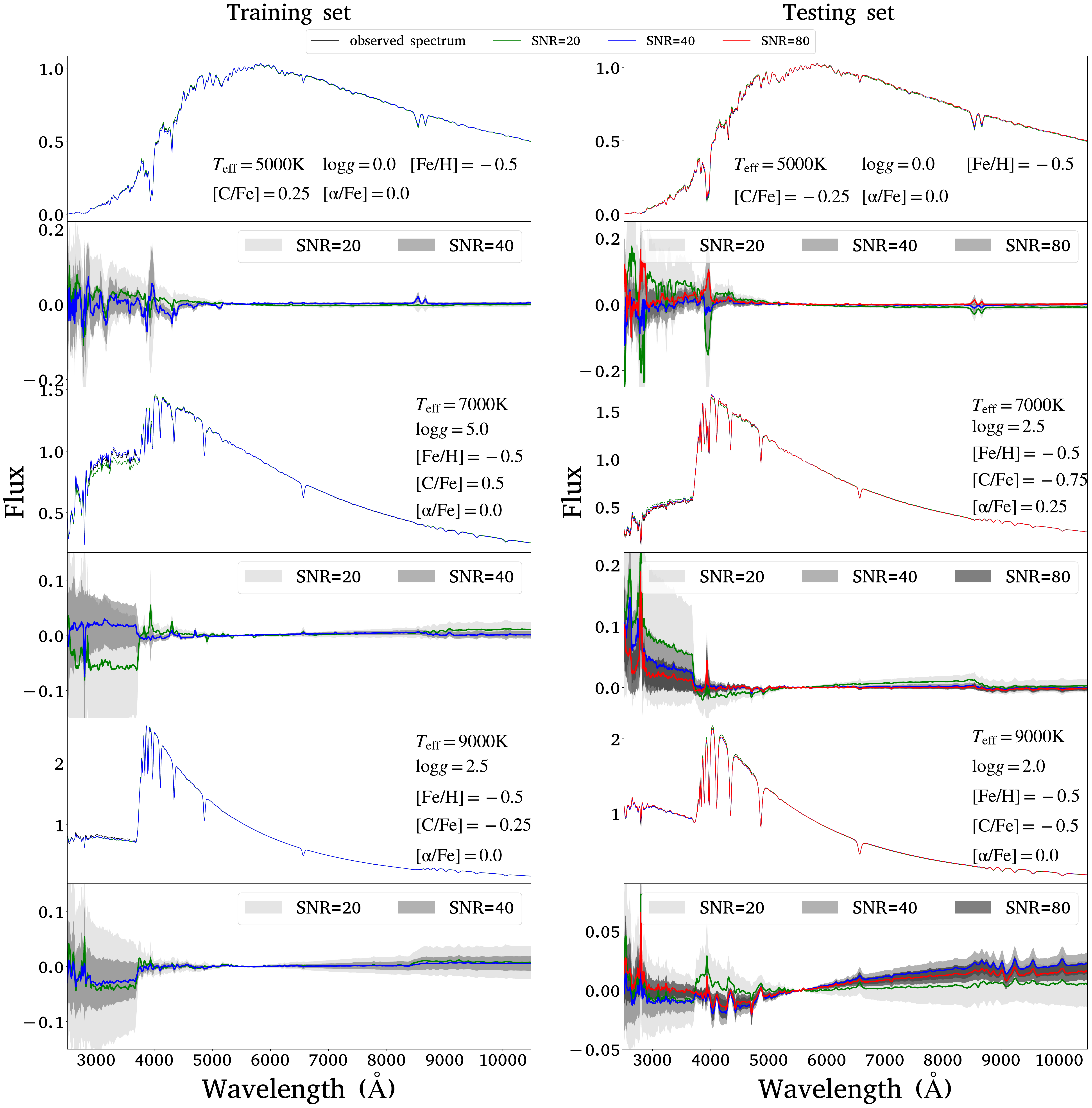}
  \caption{The SEDs from the UaRA-net and observations and their relative residuals for training set and testing set at different SNRs. A total of six representative stars of different spectral types are plotted.}
  \label{fig:spectrum}
\end{figure*}

The UaRA-net were trained using Keras 2.2.4, Tensorflow 1.12.0, CUDA 9.2, and cuDNN 7.6, running on the graphic processing units (GPUs) for acceleration. Experiments were conducted on a computer with six Intel Xeon Gold 6230 central processing units and a NVIDIA Tesla V100 SXM2 GPU. The precision and adaptability of the model were tested in this section.

\subsection{Hyperparameter setup} \label{Hyperparameter Setup}

Before pre-training and fine-tuning the UaRA-net, hyperparameters related to the architecture and optimization were set manually. Architecture-related hyperparameters include the size and number of convolutional kernels, the number of neurons and hidden layers of fully connected layers, activation function, and weight initialization method. For training CSST-like SEDs with the SEDs branch, the architecture-related hyperparameters are summarized in Table\,\ref{table:table1}. They were set based on the hyperparameters used in cost-sensitive neural networks (CSNet; \citealt{Yang2022}), deep residual shrinkage networks (DRSN; \citealt{Zhao2019}), and ResNet (\citealt{he2016}). A sub-unit consists of four convolutional layers in 3D forms of (channels, size, stride), where the three items represent the number, size, and moving stride of the convolutional kernel, respectively. For training errors of the predictive CSST-like SEDs through the uncertainty branch, three hidden layers with 16, 64, and 574 neurons, and a concatenation layer were used to extract deep representations from spectral noise density and stellar labels. The ReLU activation function and He normal initializer were used, following the setup in \citet{he2016}.

\begin{table}[ht!]
\centering
\caption{Architecture-related hyperparameters for training CSST-like SEDs and errors.}
\label{table:table1}
\begin{tabular}{cccc}
\hline
\multicolumn{2}{c}{SEDs}  & \multicolumn{2}{c}{Errors} \\
\hline
 Layers &  Output size & Layers &  Output size  \\
\hline 
  Input &  $1\times 2238\times 1$ & Input & $1\times 2238\times 1$ \\
  Cov(4,3,2) &  $4\times 1119\times 1$ &--  &--\\
  Sub-$\rm unit_1$(4,3,2) &  $4\times 560\times 1$ &-- & --\\
  Sub-$\rm unit_2$(4,3,1) &  $4\times 560\times 1$ &-- &--\\
  Sub-$\rm unit_3$(8,3,2) &  $8\times 280\times 1$ &-- &--\\
  Sub-$\rm unit_4$(8,3,1) &  $8\times 280\times 1$ &-- &--\\
  Sub-$\rm unit_5$(16,3,2) &  $16\times 140\times 1$ &-- &--\\
  Sub-$\rm unit_6$(16,3,1) &  $16\times 140\times 1$ &-- &--\\
  BN, GAP &  16 &$\rm FC'_1$ &16\\
  $\rm FC_1$ &  100 & Concat &32\\
  $\rm FC_2$ &  200 & $\rm FC'_2$ & 64\\    
  $\rm FC_3$ &  300 & $\rm FC'_3$ & 574\\
  $\rm FC_4$ &  574 &-- &--\\
\hline
\end{tabular}
\end{table}

The optimization-related hyperparameters include the optimizer, the learning rate $\eta$, the batch size of the training samples $s$, and the number of training iterations $epochs$. The adaptive moment estimation (ADAM) was used as the optimizer due to its good robustness to the initial learning rate. The learning rate $\eta$ was set to 0.001 following the recommendation of \citet{Adam}. A large batch size of $s=$2048 was set to accelerate convergence, while being limited by GPU memory. The loss function in Eq.\,(\ref{likelihood}) converged when $epoch=3000$. When fine-tuning the UaRA-net on NGSL spectra, the optimization-related hyperparameters ${\eta,s,epochs}$ were set to ${0.001,32,200}$ due to the small size of the training samples.

\subsection{Performance on predicting CSST-like spectra} \label{training and testing}
\begin{figure*}[ht!]
  \centering
  \includegraphics[width=0.75\textwidth]{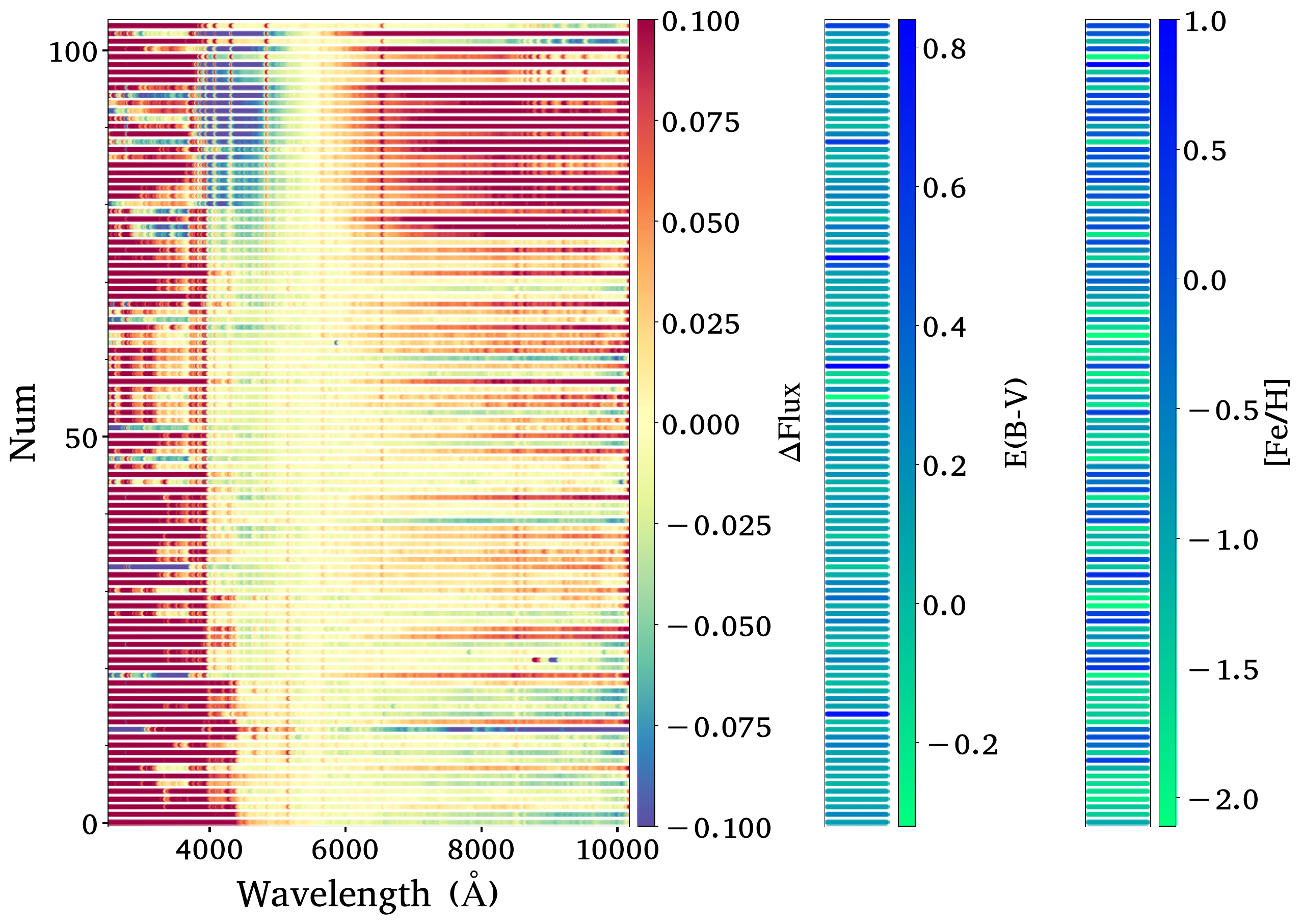}
  \caption{Relative residuals for the predicted NGSL SEDs when directly applying the pre-trained UaRA-net. Spectra are ranked by the $T_{\rm eff}$ from 5000 to 9750 K.}
  \label{fig:baseline_NGSL}
\end{figure*}

\begin{figure*}[ht!]
  \centering
  \includegraphics[width=0.75\textwidth]{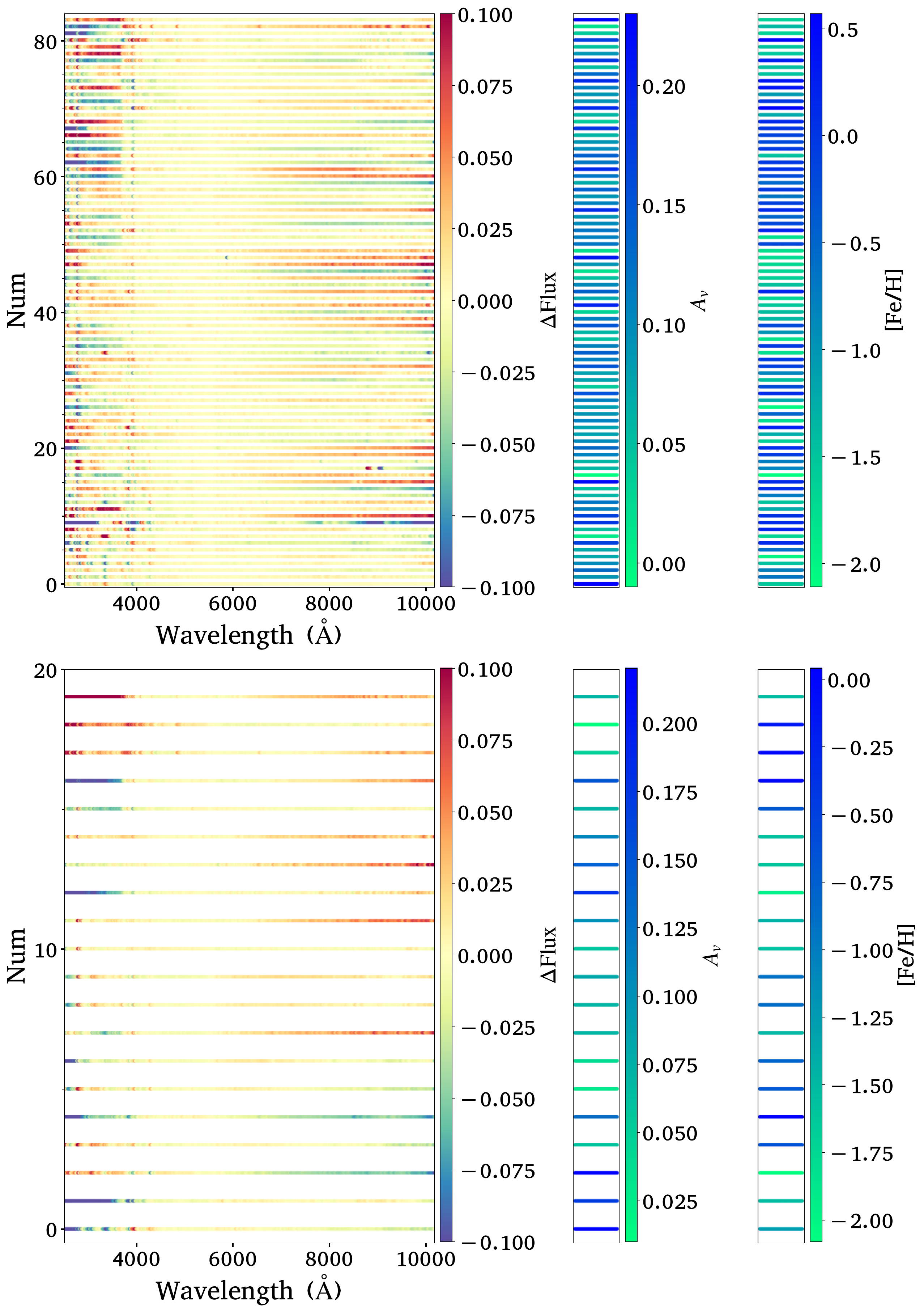}
  \caption{Relative residuals for the training (top panel) and testing (bottom panel) sets of the predicted NGSL SEDs with the fine-tuned UaRA-net.}
  \label{fig:NGSL0}
\end{figure*}

\begin{figure*}[ht!]
  \centering
  \includegraphics[width=0.75\textwidth]{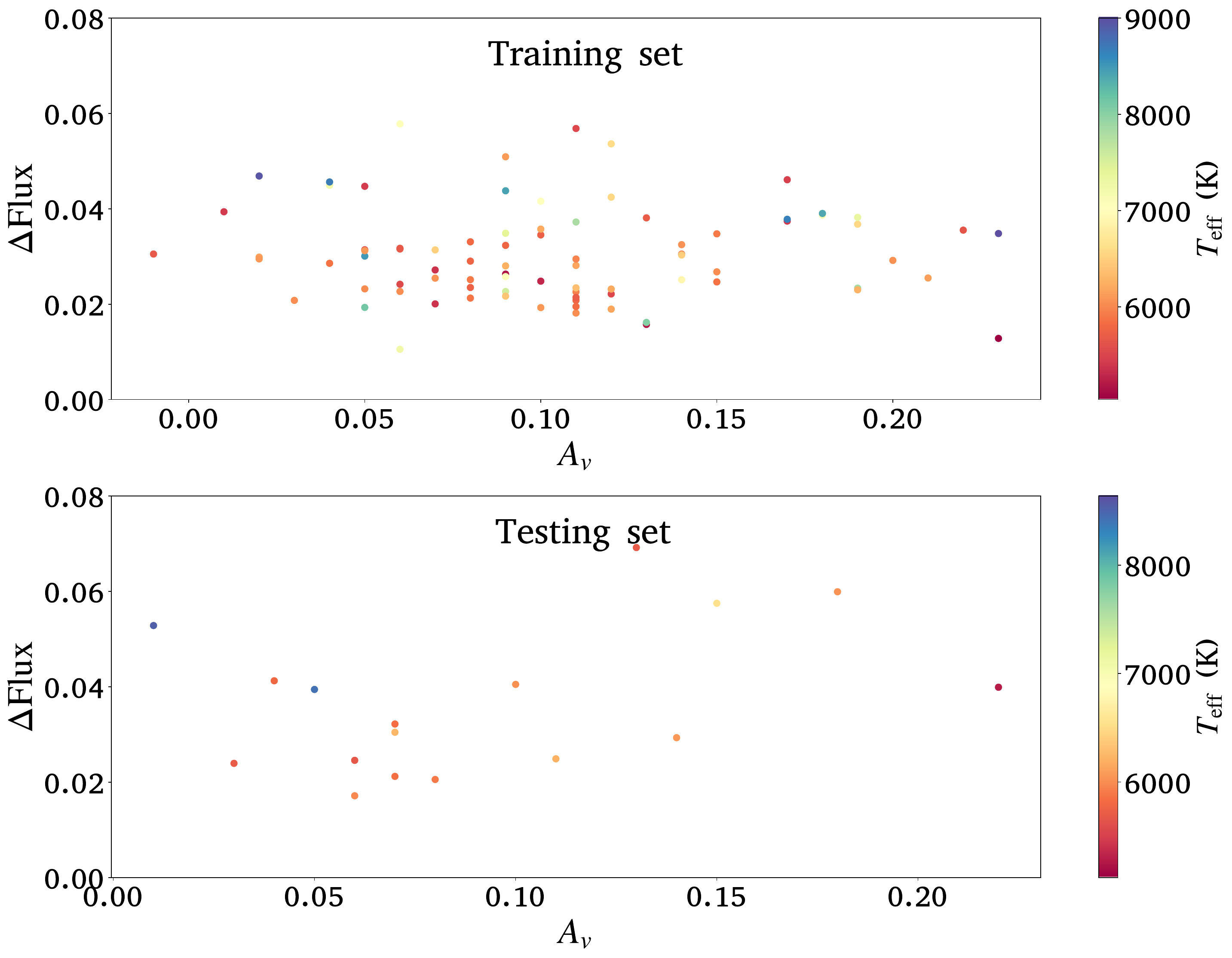}
  \caption{The $RMSAE$ of the predicted NGSL SEDs as a function of $A_V$ for the training (top panel) and testing (bottom panel) sets with the fine-tuned UaRA-net.}
  \label{fig:NGSL_ebv}
\end{figure*}

\begin{figure*}[ht!]
  \centering
  \includegraphics[width=0.75\textwidth]{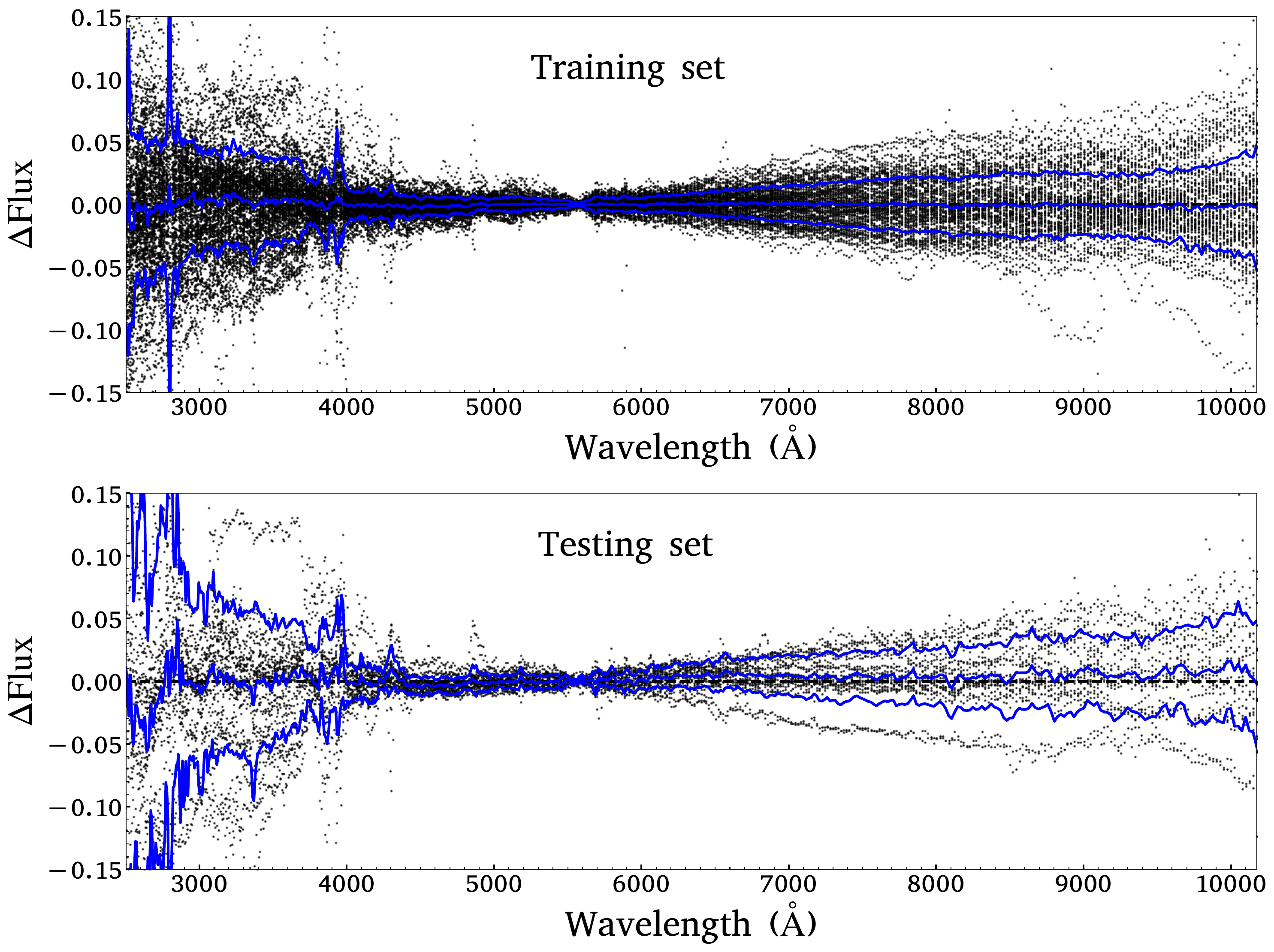}
  \caption{Relative residuals for the predicted NGSL SEDs with the fine-tuned UaRA-net. Blue lines denote the medians and the standard deviations of relative residuals as a function of wavelength. Top panel: training set. Bottom panel: testing set.}
  \label{fig:NGSL1}
\end{figure*}

\begin{figure*}[ht!]
  \centering
  \includegraphics[width=0.75\textwidth]{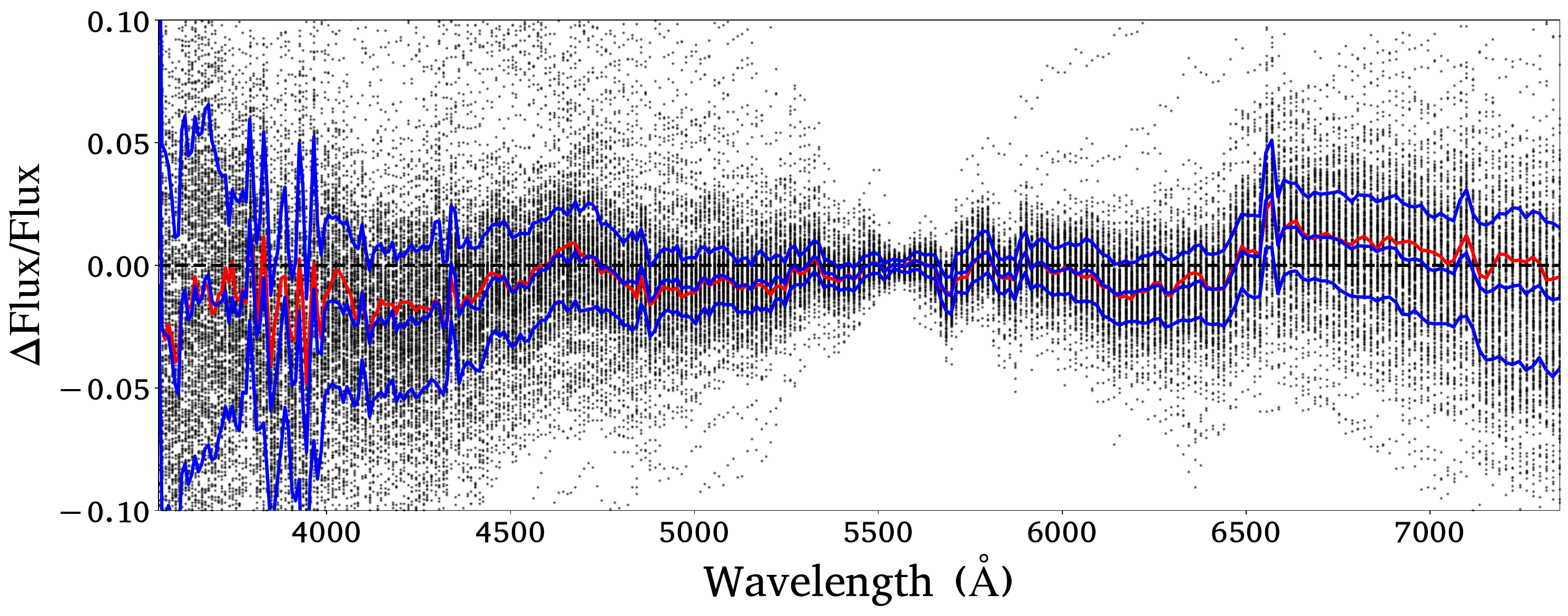}
  \caption{Similar to Fig.\,\ref{fig:NGSL1}, but for the MILES spectra. It is worth mentioning that the fine-tuned UaRA-net applied here uses the NGSL spectra rather than the MILES spectra. The red line denotes the systematic difference between the MILES spectra and NGSL spectra yielded by their common stars. The blue line and the red line match well.} 
  \label{fig:MILES1}
\end{figure*}

In this section, using the BOSZ and NGSL spectra, we evaluated the accuracy of the SEDs and uncertainties estimated from the pre-trained and fine-tuned models, respectively.

For the BOSZ spectra, we evaluated the performance of the pre-trained model on both the training and testing sets. To avoid crowding, we only show the results of the training set with $\rm SNR\in {20,40}$ and the testing set with $\rm SNR\in {20,40,80}$. Fig.\,\ref{fig:snr_u_sigma_train} shows the results of the training set. 
The medians of the relative residuals 
at both SNR $=$ 20 and 40 are close to zero, demonstrating that our predicted SEDs have a good agreement with the reference. As for the standard deviations of the relative residuals, dependencies on the $T_{\rm eff}$, wavelength, and SNRs are revealed, especially in the GU band. In this band, spectra of hot stars have higher fluxes and are less sensitive to the metallicity, leading to relatively lower errors.
For stars with $T_{\rm eff}=$ 6000 K, the typical errors in the GU, GV, and GI bands are (0.0409, 0.0024, 0.0026) and (0.0213, 0.0012, 0.0014) for the training set at SNR $=$ 20 and 40, respectively. When the $T_{\rm eff}$ increases to 8000 K, the typical errors decrease to (0.0115, 0.0016, 0.0023) and (0.0064, 0.0008, 0.0013) for the training set at SNR $=$ 20 and 40, respectively.

The results of the testing set shown in Fig.\,\ref{fig:snr_u_sigma_test} are similar to those of the training set. We note that the testing set at $\rm SNR=80$ performs well, although these samples were derived by extrapolation due to the limitation of the training set. For stars with $T_{\rm eff}=$ 6000 K, the typical errors in the GU, GV, and GI bands are, respectively, (0.0425, 0.0025, 0.0029), (0.0207, 0.0013, 0.0015) and (0.0151, 0.0011, 0.0011) for the testing set at SNR $=$ 20, 40 and 80. When the $T_{\rm eff}$ increases to 8000 K, the typical errors decrease to (0.0117, 0.0016, 0.0024), (0.0064, 0.0008, 0.0014), and (0.0052, 0.0007, 0.0012) at SNR $=$ 20, 40, and 80, respectively. The small errors demonstrate that the SEDs at $R=200$ can be well recovered from the normalized spectra at $R=2000$.

For errors estimating, we divide the samples into different $T_{\rm eff}$ bins. Fig.\,\ref{fig:sigma_train} and Fig.\,\ref{fig:sigma_test} show comparisons of the medians of predicted errors and the standard deviations of relative residuals for $T_{\rm eff}$ bins of the training set and testing set, respectively. Overall, the predicted errors are in good agreement with the true values. However, we do not expect a perfect match, as the uncertainty branch of the UaRA-net for estimating errors is an unsupervised learning algorithm. To observe the details of the relative residuals over the whole wavelength range, Fig.\,\ref{fig:spectrum} plots 6 representative spectra from the training set and the testing set. It can be seen that the pre-trained model performs well, and the predicted errors have a strong negative correlation with the SNRs, as expected.

For the NGSL spectra, we first evaluated the performance of the pre-trained model. Fig.\,\ref{fig:baseline_NGSL} shows relative residuals of the pre-trained model.  It can be seen that the relative residuals have a systematic trend with $T_{\rm eff}$, which is caused by the systematic discrepancies between the BOSZ theoretical spectra and the NGSL empirical spectra. The discrepancies mainly come from systematic uncertainties in the BOSZ theoretical spectra, particularly in the UV. Then, we evaluated the performance of the fine-tuned model on both the training and testing set, as shown in Fig.\,\ref{fig:NGSL0}. The good consistency demonstrates that the fine-tuned model is now capable of adapting spectra from BOSZ to NGSL. Fig.\,\ref{fig:NGSL_ebv} plots the difference between the SEDs from the fine-tuned model and observations as a function of $A_V$. No obvious dependence neither $A_V$ nor $T_{\rm eff}$ is observed in both the training and testing sets. Fig.\,\ref{fig:NGSL1} shows the distribution of relative residuals as a function of wavelength for both the training and testing sets. The relative residuals show no systematic patterns with wavelength.

\subsection{Verification with MILES spectra}
To demonstrate the reliability of the proposed method, we applied the fine-tuned UaRA-net to MILES spectra and obtained their predicted SEDs. Fig.\,\ref{fig:MILES1} shows the relative residuals of the 265 MILES spectra as a function of wavelength. The typical relative residuals over the whole wavelength range are less than 0.02. The trend of the relative residuals with wavelength (the red line) is consistent with the systematic difference between the NGSL spectra and MILES spectra (the black line). 
Note that the discrepancy at  $\lambda > 6700\,\rm \AA$ is attributed to the systematic errors (imperfect correction of second-order contamination; \citealt{Sanchez-Blazquez2006}) in the MILES.
The errors at $\lambda \sim $3600, 4500, 5000, 6000 and 7000$\rm \AA$ are 0.070, 0.022, 0.013, 0.010, and 0.022, respectively. The small errors suggest that the proposed SEDs estimation pipeline can be generally applied to observed spectra.


\section{Summary and Future Perspective} 
\label{summary} 

In this work, we designed and trained an uncertainty-aware residual attention network, the UaRA-net, to convert normalized spectra across the wavelength range of 4000--7000$\rm \AA$ into SEDs spanning 2500--10500$\rm \AA$. The UaRA-net incorporates the effects of stellar parameters and SNR to estimate precise SEDs and their associated uncertainties. We first pre-trained a baseline model using degraded BOSZ spectra, which simulate the CSST stellar spectra, and then further fine-tuned the model with other CSST-like spectra, such as the NGSL and MILES spectra. We found that the precision of the predicted SEDs depends on $T_{\rm eff}$, wavelength and SNR. At $\rm SNR=20$, the maximum offsets and errors in the GU, GV, and GI bands are (0.072, 0.198), (0.007, 0.032), and (0.005,0.014), respectively. Using millions of stellar spectra from LAMOST, this method can be applied to the flux calibration of CSST and other space-based spectroscopic surveys in the future. 

To investigate the precision of flux calibration by our method, following the procedure outlined in \cite{Yuan2021}, we randomly selected 400 stars with $\rm SNR>20$ and $G>14$ from LAMOST DR7. Assuming that these stars would be observed by the CSST, we estimated calibration errors of the UaRA-net using Monte Carlo simulations. We fitted the differences in flux among the 400 spectra using a 2nd-order 2D polynomial function to generate the flat-field cube.
We achieved typical flux calibration precision of 0.005, 0.0003, and 0.0005 for the GU, GV, and GI bands, respectively. Note that these numbers 
are too optimistic and only valid under ideal conditions.

The proposed UaRA-net assumes that the epistemic
uncertainty is far smaller than the aleatoric uncertainty of the model. 
However, when the number of training samples is small, the epistemic uncertainty becomes significant. Therefore, the performance of the UaRA-net can be improved by fine-tuning with more well-calibrated empirical spectra. Also, the UaRA-net can be directly trained with a number of well-calibrated CSST spectra in the future. 

We assumed that we know well the flux at 5526--5586$\rm \AA$ of our LAMOST stars. However, it is important to note that these flux measurements have not been subject to absolute flux calibration. 
In the future, we plan to determine the absolute flux at  5526--5586 $\rm \AA$ using Gaia photometry and XP spectra.  In addition, the flux in the UV band is sensitive to metallicity, leading to large uncertainties in the UV.  We are going to improve the correction performance in the UV by incorporating white dwarfs in the future.

We assumed that the resolution ($R=200$) of the CSST slitless spectra is uniform across the entire wavelength range. However, the actual resolution varies slightly across the GU, GV, and GI bands, which needs to be taken into account in future. In the current method, we also ignored the effect 
of dust reddening. Such effect should be taken into account in the future.

Gaia DR3 provides high-quality low-resolution ($R\sim50$) BP and RP spectra covering the wavelength range 3300--10500 $\rm \AA$. After a comprehensive correction of systematic errors in the flux calibration \citep{Huang2024}, it provides an alternative dataset for the UaRA-net to deliver a large number of flux-standard stars.

\vspace{7mm} \noindent {\bf Acknowledgments}
This work is supported by the National Natural Science Foundation of China through the projects NSFC 12173007, 11603002, 12222301, National Key Basic R \& D Program of China via 2019YFA0405500, 
and Beijing Normal University grant No. 310232102. 
We acknowledge the science research grants from the China Manned Space Project with NO. CMS-CSST-2021-A08 and CMS-CSST-2021-A09.
Guoshoujing Telescope (the Large Sky Area Multi-Object Fiber Spectroscopic Telescope LAMOST) is a National Major Scientific Project built by the Chinese Academy of Sciences. Funding for the project has been provided by the National Development and Reform Commission. LAMOST is operated and managed by the National Astronomical Observatories, Chinese Academy of Sciences. 
This work has made use of data from the European Space Agency (ESA) mission {\it Gaia} (https://www.cosmos.esa.int/gaia), processed by the {\it Gaia} Data Processing and Analysis
Consortium (DPAC, https://www.cosmos.esa.int/
web/gaia/dpac/ consortium). Funding for the DPAC has been provided by national institutions, in particular the institutions participating in the {\it Gaia} Multilateral Agreement.
Data resources are supported by China National Astronomical Data Center (NADC) and Chinese Virtual Observatory (China-VO). This work is supported by Astronomical Big Data Joint Research Center, co-founded by National Astronomical Observatories, Chinese Academy of Sciences and Alibaba Cloud.


\appendix

\section{Gaussian process regression}\label{GPR}
Regression models based on Gaussian processes (GPR), a full probabilistic Bayesian approach, are used in this work to establish the relationship between stellar parameters ($T_{\rm eff}$, $\log g$, and [Fe/H]) and the SEDs. The stellar parameters $X$ are standardized into $(X-\mu)/\sigma$ by z-score normalization, where $\mu$ and $\sigma$ are the mean and standard deviation of the original stellar parameters, respectively.
We assume that the SED $Y$ is generated by the function $f(X)$ plus additive noise $\epsilon \in \mathcal N(0,\sigma_n^2)$, where $f(X)$ is drawn from the Gaussian process on $X$ specified by the covariance function $K(X,X^{'})$. 
The process can be described as follows
\begin{gather}
    Y=f(X)+\epsilon\\
    f(X)\sim \mathcal G \mathcal P(m(X),K(X,X^{'}))
\end{gather}
where 
\begin{gather}
    m(X)=\mathbb{E}(f(X))\\
    K(X,X^{'})=\mathbb{E}[(f(X)-m(X))(f(X^{'})-m(X^{'}))]
\end{gather}
In this work, the covariance function $K(X,X^{'})$ are specified to be squared exponential kernel plus noise-component of the signal as
\begin{equation}
    K(X,X^{'})=\sigma _{f}^{2}\exp \left(-\frac{\left \| X-X^{'}\right \|^{2}}{2l^2}\right)
\end{equation}
where $l$ is an isotropic length-scale parameter, and $\sigma_f$ is the standard deviation of the signal.
The Gaussian process that target values $Y$ is drawn from can be defined as
\begin{equation}
    Y\sim \mathcal G \mathcal P(m(X),K(X,X^{'})+\sigma_n^{2}\delta_{XX^{'}})
\end{equation}
where $\delta$ is the Kronecker delta function.
According to the definition of Gaussian process, given the training set $(X,Y)$ and testing set $(X_{*},Y_{*})$, the conditional distribution $Y_{*}|Y,X,X_{*}\sim \mathcal{N}(\mu, \Sigma)$ is normal, where
\begin{gather}
    \mu = K(X,X_{*})(K(X,X)+\sigma_{n}^{2}\textbf{I})^{-1}Y\\
    \Sigma = K(X_{*},X_{*})-\sigma_{n}^{2}-(K(X,X)+\sigma_{n}^{2}\textbf{I})^{-1}K(X,X_{*})
\end{gather}
Parameters of the GPR model, $\theta=\{ \sigma_f,\sigma_n,l \}$, are estimated by maximizing the marginal likelihood
\begin{equation}
    P(Y|X)=\int P(Y|f,X)P(f|X)df
\end{equation}
This leads to the minimization of
the negative log-posterior
\begin{equation}
    \begin{aligned}
    -\log P(Y|X,\theta)=&\frac{1}{2}Y^{T}(K+\sigma_n^{2}\textbf{I})^{-1}Y+\frac{1}{2}\log\left | K+\sigma_n^{2}\textbf{I}\right |\\&-\log p(\sigma_n^{2})-\log p(k)
    \end{aligned}
\end{equation}

\begin{figure*}[ht!]
  \centering
  \includegraphics[width=0.75\textwidth]{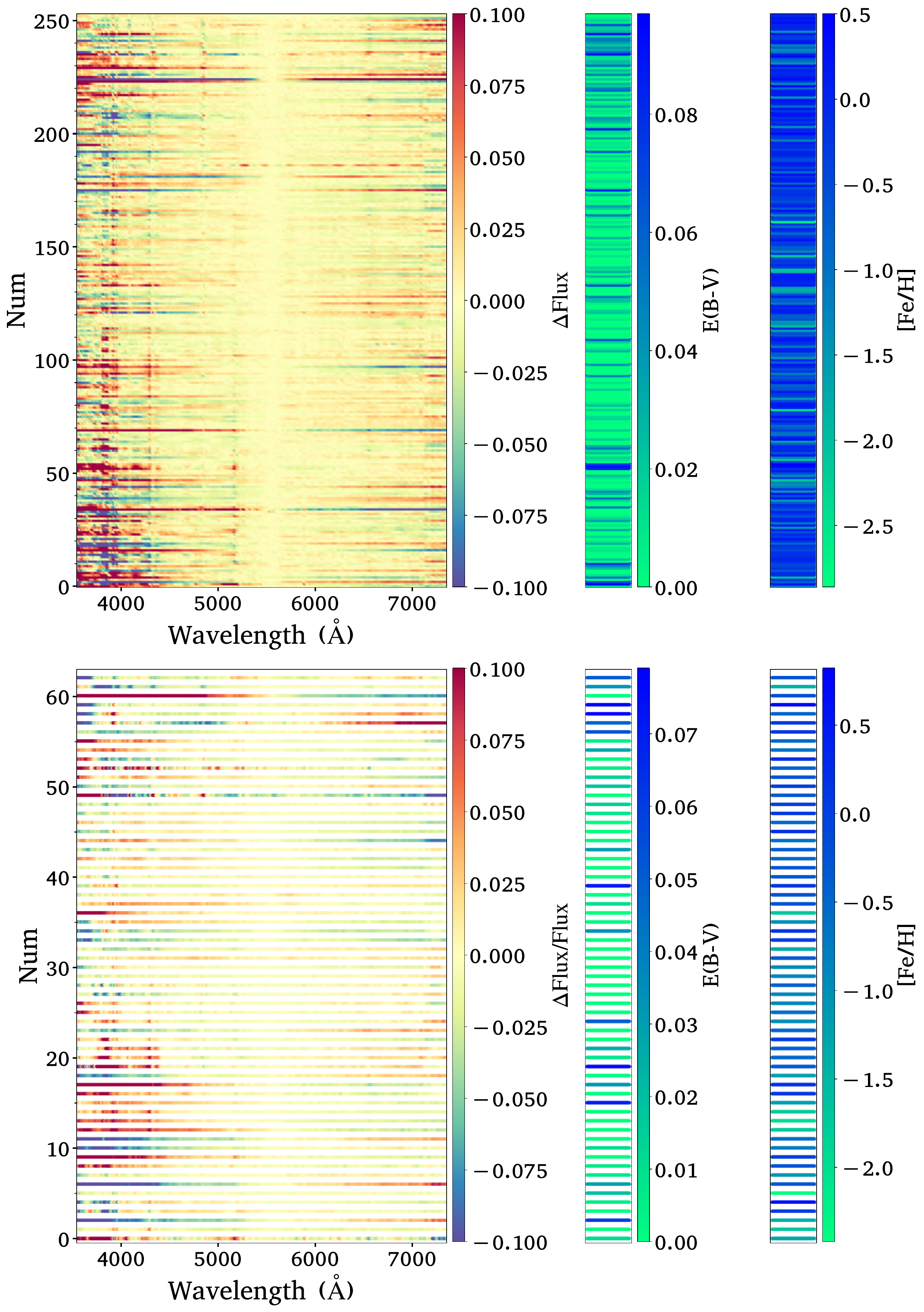}
  \caption{Relative residuals of the GPR training (top panel) and testing (bottom panel) sets for MILES. }
  \label{fig:MILES_GPR}
\end{figure*}

\begin{figure*}[ht!]
  \centering
  \includegraphics[width=0.75\textwidth]{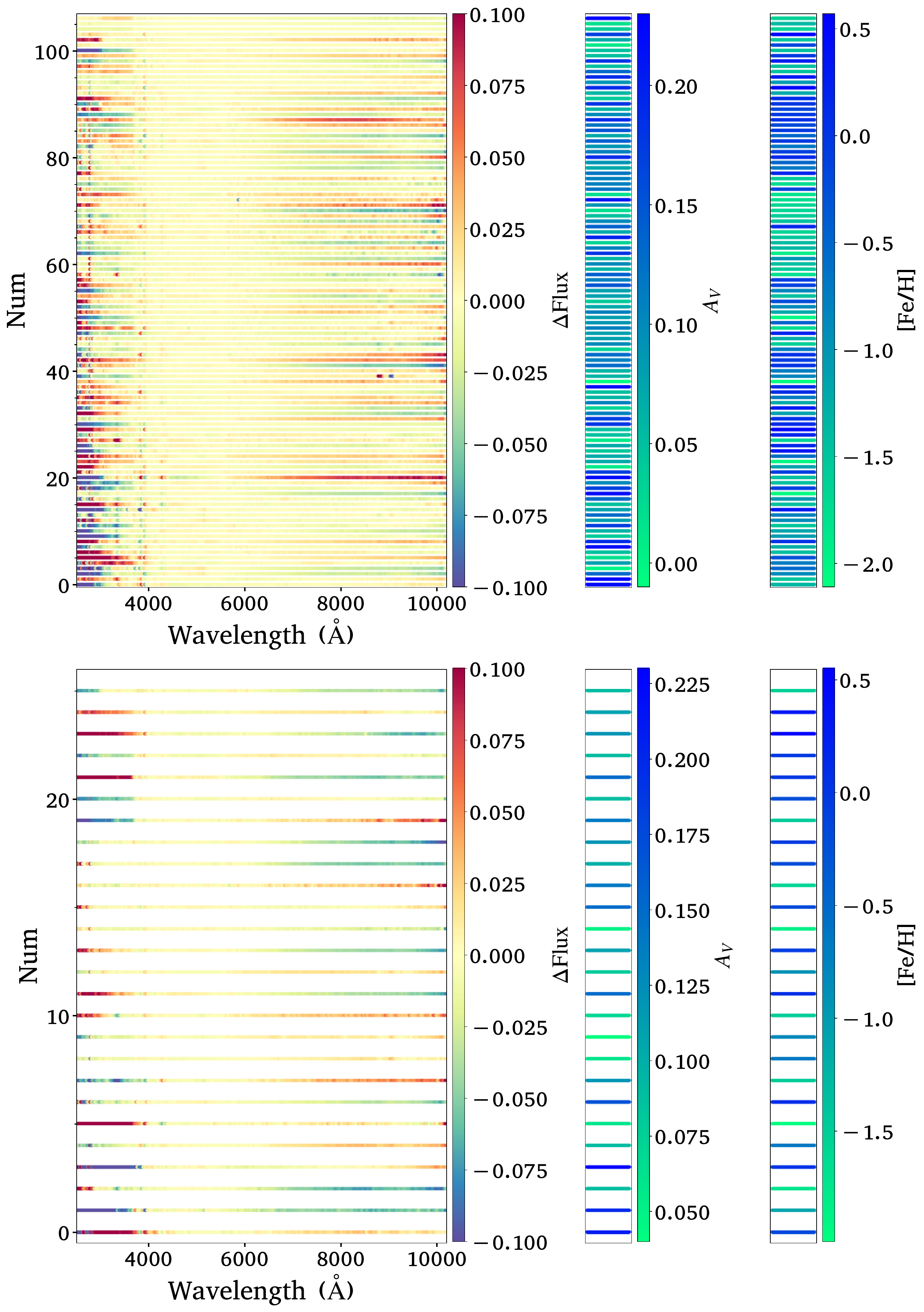}
  \caption{Similar to Fig.\,\ref{fig:MILES_GPR}, but for NGSL. }
  \label{fig:NGSL_GPR}
\end{figure*}

\begin{figure*}[ht!]
  \centering
  \includegraphics[width=0.75\textwidth]{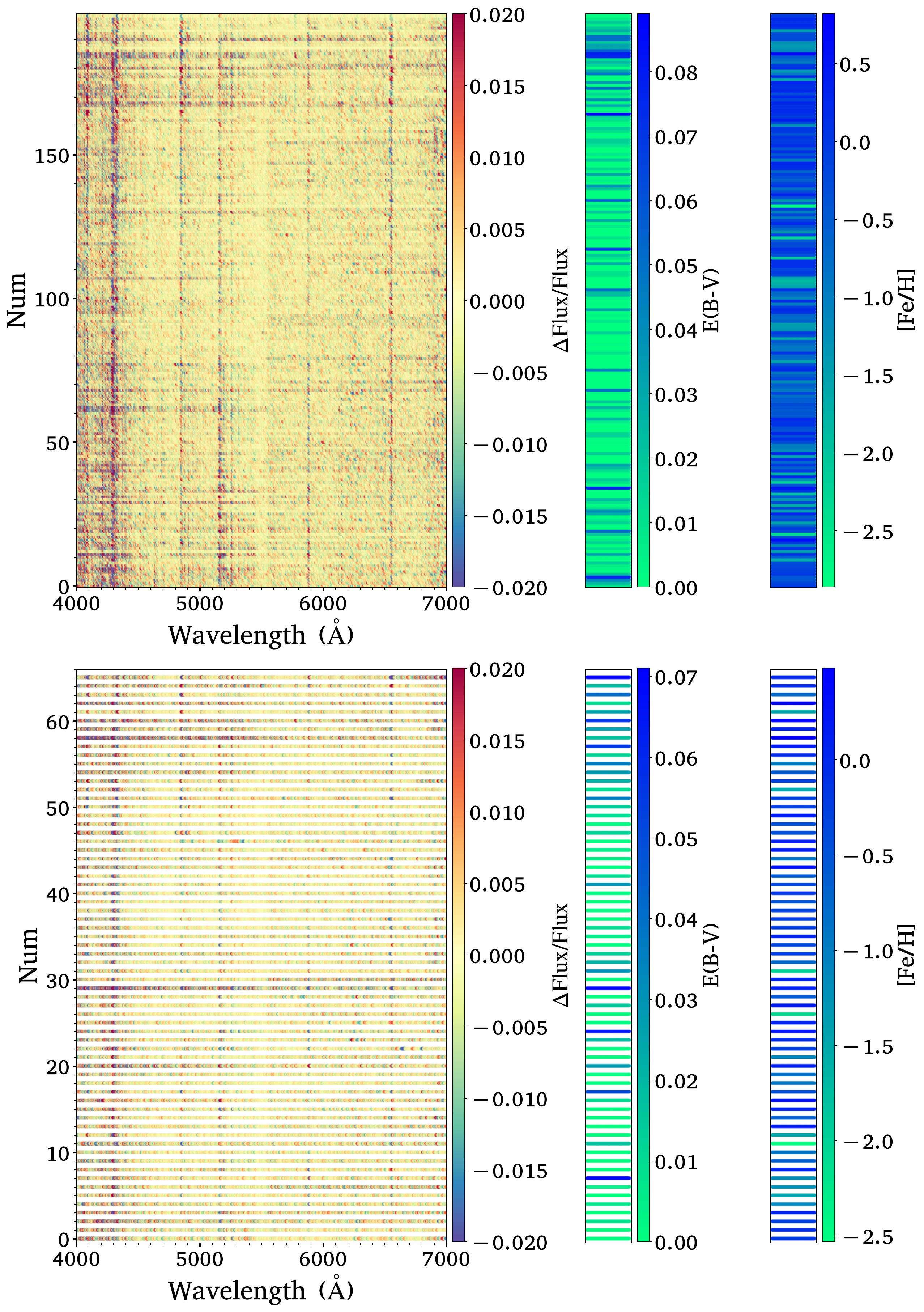}
  \caption{Relative residuals for training (top panel) and testing (bottom panel) sets of the MILES results at $R=2000$ using GPR.}
  \label{fig:MILES_GPR_R2000}
\end{figure*}

\begin{figure*}[ht!] 
  \centering
  \includegraphics[width=0.75\textwidth]{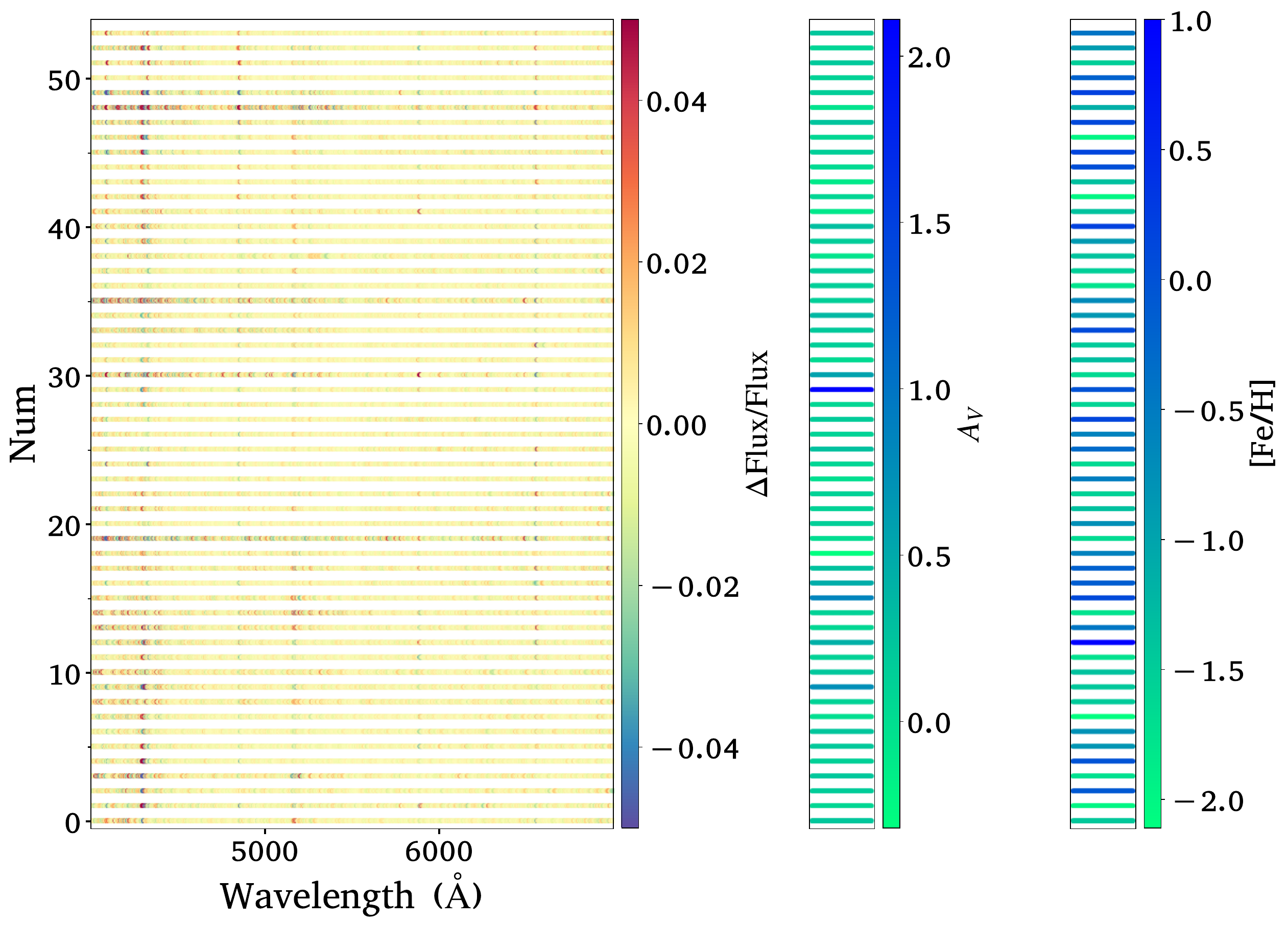}
  \caption{Relative residuals between the generated NGSL-based spectra and the MILES spectra of the common sources.}
  \label{fig:cmp_NGSL_GPR_R2000}
\end{figure*}

\section{Data selection strategy for MILES and NGSL spectra}\label{data selection}

Two GPR models were trained in Section\,\ref{GPR} to establish the relationship between the stellar parameters and the normalized spectra at $R=200$ for the MILES and NGLS spectra, respectively. We selected 316 stars with $5000 \leq T_{\rm eff} \leq 9750$ K and $E(B-V)<0.1$ mag from MILES library, and 136 stars with $5000 \leq T_{\rm eff} \leq 9750$ K, $-0.05<A_V<0.1$ mag, err($T_{\rm eff}$) $<$ 500 K, err($\log g$) $<$ 0.5 dex, and err([Fe/H]) $<$ 0.5 dex from NGSL library for training and testing the above two models, respectively. Both sets of selected stars were randomly divided into training and testing set in the ratio of 4:1. The trained hyper-parameters $\theta$ of the GPR model for the MILES and NGSL spectra were $\{ 0.883,0.001,4.27 \}$ and $\{ 0.613,0.0003,2.57 \}$, respectively.

Fig.\,\ref{fig:MILES_GPR} and Fig.\,\ref{fig:NGSL_GPR} show relative residuals of the GPR results in the training and testing sets for the MILES and the NGSL, respectively. It can be seen that the relative residuals of most spectra are small, expect for several outliers.
Finally, 265 MILES spectra with $RMSAE<0.05$ and 115 NGSL spectra with $RMSAE<0.068$ were selected.

\section{NGSL-based spectral generation}\label{NGSL}

Taking the stellar parameters ($T_{\rm eff}$, $\log g$, and [Fe/H]) as the inputs and the normalized spectra of $R=2000$ at $4000<\lambda<7000\AA$ as the outputs, we trained a GPR model on 265 selected MILES spectra from Section\,\ref{data selection}. Fig.\,\ref{fig:MILES_GPR_R2000} shows relative residuals of the training and testing sets for the MILES spectra. Overall, results from the GPR model are in good agreement with the observations. Then, the trained model was applied to the NGSL spectra to obtain the NGSL-based spectra at $R=2000$. Fig.\,\ref{fig:cmp_NGSL_GPR_R2000} shows relative residuals of 53 stars in common between MILES and NGSL-based spectra at $R=2000$. Good agreement between the MILES and NGSL demonstrates that the NGSL spectra of $R=2000$ at $4000<\lambda<7000\rm \AA$ are reliable.

\end{document}